\documentclass[aps,prb,amsmath,twocolumn,amssymb,floatfixng,showpacs,
superscriptaddress,footinbib]{revtex4-1}
\pdfoutput=1
\usepackage[dvips]{graphics}
\usepackage{bm}
\usepackage{wrapfig}
\usepackage{float}
\usepackage{epsfig}
\usepackage{enumerate}
\usepackage{color}
\usepackage{graphicx}
\usepackage{suffix}
\usepackage{mathtools}
\DeclarePairedDelimiterX\MeijerM[3]{\lparen}{\rparen}%
{\begin{smallmatrix}#1 \\ #2\end{smallmatrix}\delimsize\vert\,#3}
\newcommand\MeijerG[8][]{%
  G^{\,#2,#3}_{#4,#5}\MeijerM[#1]{#6}{#7}{#8}}

\WithSuffix\newcommand\MeijerG*[7]{%
  G^{\,#1,#2}_{#3,#4}\MeijerM*{#5}{#6}{#7}}

\usepackage[colorlinks=true,linktoc=page,linkcolor=magenta,citecolor=blue]{hyperref}

\newcommand{\etal}{{\it et al.~}}
\newcommand{\ie}{{\it i.e., }}
\newcommand{\eg}{{\it e.g. }}

\newcommand\bea{\begin{eqnarray}}
\newcommand\eea{\end{eqnarray}}
\newcommand\beq{\begin{equation}}  
\newcommand\eeq{\end{equation}}

\newcommand{\non}{\nonumber}  
\begin{document} 
\title{Fingerprints of tilted Dirac cones on the RKKY exchange interaction in 8-Pmmn Borophene} 

\author{Ganesh C. Paul}
\email{ganeshpaul@iopb.res.in}
\affiliation{Institute of Physics, Sachivalaya Marg, Bhubaneswar-751005, India}
\affiliation{Homi Bhabha National Institute, Training School Complex, Anushakti Nagar, Mumbai 400085, India}
\author{SK Firoz Islam}
\email{firoz@iopb.res.in}
\affiliation{Institute of Physics, Sachivalaya Marg, Bhubaneswar-751005, India}
%\affiliation{Department of Physics, Beijing Computational Science Research Center, Beijing 100193, China}
\author{Arijit Saha}
\email{arijit@iopb.res.in}
\affiliation{Institute of Physics, Sachivalaya Marg, Bhubaneswar-751005, India}
\affiliation{Homi Bhabha National Institute, Training School Complex, Anushakti Nagar, Mumbai 400085, India}

\begin{abstract}
We theoretically investigate the indirect signatures of the tilted anisotropic Dirac cones  
on Ruderman-Kittel-Kasuya-Yosida (RKKY) exchange interaction in a two dimensional 
polymorph of boron atoms. Unlike the case of isotropic non-tilted Dirac material-graphene, 
here we observe that the tilting of the Dirac cones exhibits a significant impact on the RKKY 
exchange interaction in terms of the suppression of oscillation frequency. The reason can be
attributed to the behavior of the Fermi level and the corresponding density of states with
respect to the tilting parameter. The direct measurement of the period of the RKKY interaction
can thus be a possible probe of the tilt parameter associated with the tilted Dirac cones. We also obtain the
direction dependent analytical expressions of the RKKY exchange interaction, in terms of Meijer G-function. 
However, the effects due to tilting of the Dirac cones on the RKKY interaction depend on the spatial alignments
of the two magnetic impurities with respect to the direction of tilting.
\end{abstract}

\maketitle
%----------------------------------------------------------------------
\section{Introduction}{\label{sec:I}}
%----------------------------------------------------------------------
With the advent of graphene~\cite{novoselov2005two, RevModPhys.81.109}, monolayer materials with similar band 
structure have been under active consideration from the theoretical and experimental research point of view
owing to their future application in nanoelectronics. Along this direction, a thin layer of boron atoms called
borophene is the latest addition to the famility of Dirac materials. Having one less electron than carbon, boron's honeycomb structure is unstable. 
However, it can be stabilized by adding extra boron atoms in the honeycomb lattice. First principle calculations have predicted that depending on the arrangements of 
the extra boron atoms, various stable monolayer-boron structures, such as $\alpha$ sheet, $\beta$ sheet, are possible~\cite{tang2010first,wu2012two,penev2012polymorphism}. 
In recent times, Mannix \etal reported a stable striped phase and a metastable homogeneous phase in two-dimensional (2D) boron silver substrate~\cite{mannix2015synthesis} 
while Feng \etal has experimentally confirmed the presence of Dirac fermions in this phase, named $\beta_{12}$ sheet~\cite{feng2017dirac}. The 8-Pmmn borophene is one of the most 
recent 2D polymorph of boron atoms, which is predicted to host tilted Dirac cones~\cite{zhou2014semimetallic}.
 
The RKKY exchange interaction~\cite{ruderman1954indirect,kasuya1956electrical,yosida1957k}
is an indirect exchange coupling between two magnetic impurities being mediated by the conduction
electrons of the host material. Since the RKKY exchange interaction is directly related to the
susceptibility of the host material, it can be used as a probe to an electronic system.
In recemt times, the physics of RKKY exchange interaction has been widely investigated in a variety of
Dirac materials such as graphene~\cite{saremi2007rkky,brey2007diluted,black2010rkky,
kogan2011rkky,sherafati2011rkky,sherafati2011analytical,uchoa2011kondo}, silicene~\cite{zare2016topological},
phosphorene~\cite{islam2018probing,duan2017anisotropic},
topological insulator~\cite{liu2009magnetic,biswas2010impurity,
zhu2011electrically,PhysRevB.90.125443}, Dirac semimetal~\cite{chang2015rkky,kaladzhyan2018rkky} etc.

Formally, the magnitude of the RKKY exchange interaction is anticipated to be severely
influenced by the position of the Fermi level and the corresponding density of states (DOS)
in any host material. In this context, the tilting of the Dirac cone can modify DOS as well
as Fermi level~\cite{zare2016topological,shiranzaei2017effect} near the Dirac point in anisotropic 
Dirac materials. Hence, the features of the RKKY interaction can carry the signatures of 
the tilting nature of the Dirac cones. This issue has not been addressed so far to the best of our knowledge.

In this work, we explore the consequences of the tilted and anisotropic Dirac cones on
the RKKY exchange interaction, considering $8$-Pmmn borophene as a host material. 
We obtain semi-analytical results of the RKKY exchange interaction for two 
different spatial separation of the magnetic impurities: two impuries are located 
perpendicular to the tilt axis and parallel to the tilt axis. For the former case,
interference between the Dirac fermions from different valleys do not contribute to the oscillation frequency and the
period of oscillation increases as one enhances the value of the tilt parameter.
This change of oscillation frequency may be a possible way to probe
the degree of tilting of the Dirac cone present in anisotropic Dirac materials such as 
$8$-Pmmn borophene. On the other hand, for the separation of the
two impurities being along the tilt axis (along the $y$ axis), interference among
the Dirac cones plays a dominant role in determining the period of oscillation and tilting parameter 
exhibits negligible effect on the corresponding period. We also demonstrate the role of tilted and anistropic Dirac cone 
on Fermi level which in turn influences the RKKY exchange interaction. Behavior of RKKY exchange interaction
is also investigated numerically for two spatially separated magnetic impurities in the $x$-$y$
plane of the 2D borophene sheet.

The remainder of the paper is structured as follows. In Sec.~\ref{sec:II}, we describe the model
Hamiltonian for our setup and present a brief outline of the Green’s function formalism to obtain the 
RKKY exchange interaction. The behavior of RKKY interaction, as a function of distance between the two
magnetic impurities as well as tilting parameter, for different alignements of the impurities is presented in 
Sec.~\ref{sec:III}. Finally, we summarize our results and conclude in Sec.~\ref{sec:IV}.

%----------------------------------------------------------------------------------------------------------------------------
%---------------------------------------------------------------------------------------------------------------------------- 
\begin{figure}[!thpb]
\centering
\includegraphics[width=0.9 \linewidth]{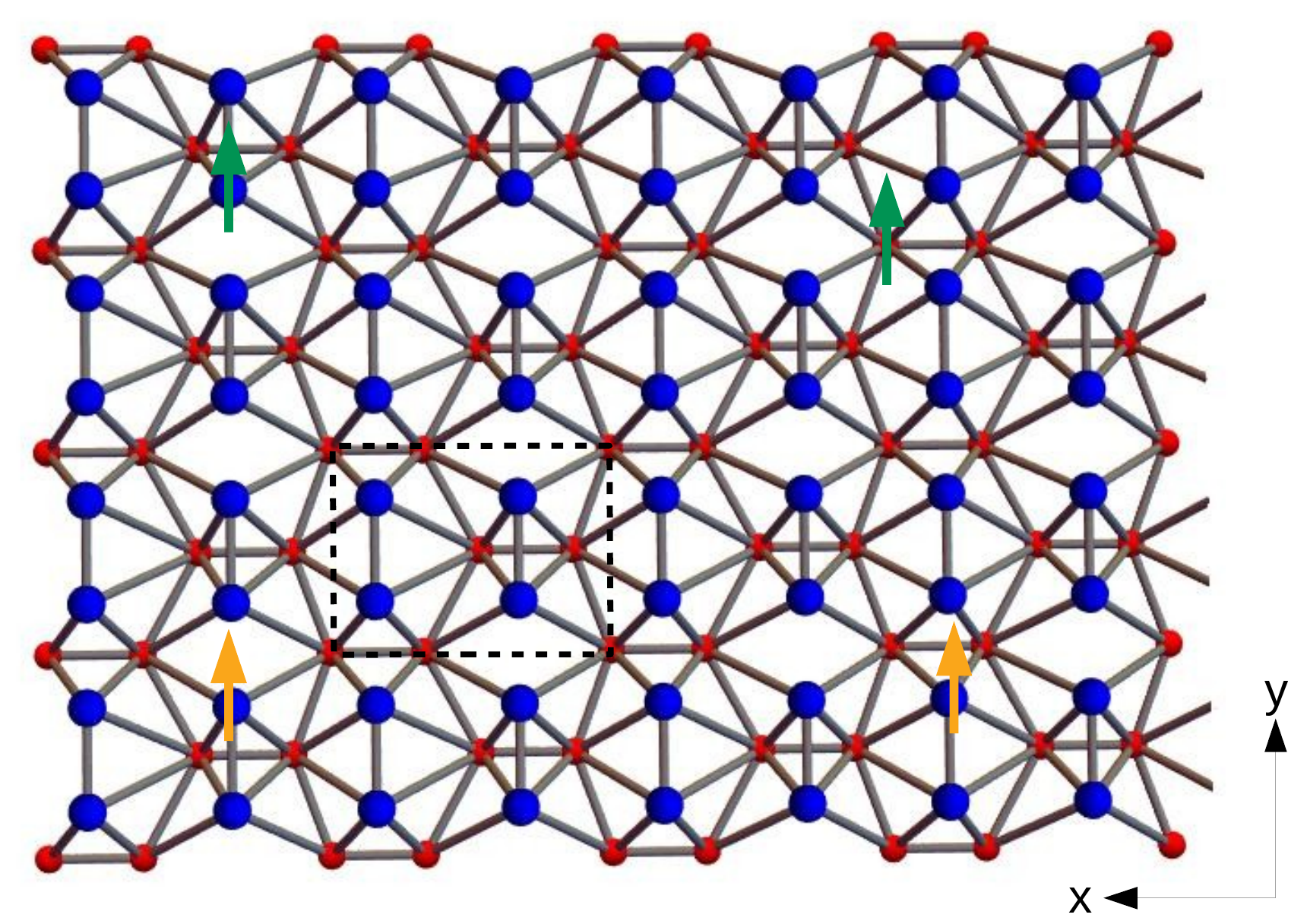}
\caption{(Color online) Schematic of the lattice structure of $8$-Pmmn borophene. Big blue (dark gray)
circles and small red (light gray) circles distinct two types of nonequivalent atoms $B_R$ (ridge atom)
and $B_I$ (inner atom) respectively. The unit cell, comprising of $8$ atoms, is shown by dashed black
rectangle. Magnetic impurities are schematically shown by golden (light gray) and dark green
(dark gray) arrows. Bottom (golden) arrows are aligned perpendicular to the tilt axis ($x$ axis) while the left arrows
(golden and dark green) are located along the tilt axis ($y$ axis) inside the 2D sheet.}
\label{model}
\end{figure}
%-----------------------------------------------------------------------------------------------------------------------------
%-----------------------------------------------------------------------------------------------------------------------------

%----------------------------------------------------------------------
\section{Model and Method}{\label{sec:II}}
%----------------------------------------------------------------------
The $2$D lattice structure of $8$-Pmmn borophene  consists of two kinds of nonequivalent atoms,
namely, ridge atoms ($B_R$) and inner atoms ($B_I$)~\cite{zhou2014semimetallic}. The corresponding lattice structure
is characterized by the lattice parameters: $a= 4.51~\rm\AA$ and 
$b=3.25~\rm\AA$~\cite{lopez2016electronic,zabolotskiy2016strain} and schematically shown in 
Fig.~\ref{model}. Here the unit cell is depicted by the dashed black rectangle (see Fig.~\ref{model}).

Although the complete tight-binding model for $8$-Pmmn borophene is quite tedious, a low energy
effective Hamiltonian has been recently put forwarded by Zabolotskiy and
Lozovic~\cite{zabolotskiy2016strain} using symmetry consideration and has been
successively implemented in a series of theoretical works exploring anisotropic
plasmon~\cite{sadhukhan2017anisotropic}, valley dependent integer quantum Hall effects~\cite{Islam_2018},
Weiss oscillation~\cite{islam2017signature},
Klein tunneling~\cite{zhang2018oblique}, optical conductivity~\cite{verma2017drude} etc.
in this $2$D material. Hence, we begin with a general low energy two band effective
Hamiltonian for $2$D Dirac materials associated with anisotropic Dirac cone, which can be written
as (near the Dirac points $\bf{k}=\pm \bf{k_D}$)
\begin{eqnarray}
&&H_D=\xi(v_x \sigma_x q_x +v_y \sigma_y q_y+ v_t \sigma_0 q_y) \ ,
\label{Ham}
\end{eqnarray}
where $\sigma_{x}, \sigma_{y}$ are the Pauli matrices in the atomic basis and $\sigma_{0}$ is a unit matrix.
We have chosen $\hbar=1$. Here, $\xi=\pm 1$ is the valley index and $v_{i}$ ($i=x,y$) corresponds to the velocities along $i$$^{\rm th}$ direction, 
while $v_t$ denotes the velocity scale associated with the tilted Dirac cones. Note that the tilting is along the $y$-direction.
The different velocity parameters are given by $\{ v_x,v_y\}=\{0.86,0.69\}$ and 
$v_t=0.32$ in units of $10^6$ m/sec~\cite{lopez2016electronic,zabolotskiy2016strain}.
The above Hamiltonian can be further written as
\begin{equation}
 H_D=\xi[ v_t \sigma_0 q_y+ v_F(\sigma_x\tilde{q}_x+\sigma_y\tilde{q}_y)]\ ,
\end{equation}
where $v_F=\sqrt{v_xv_y}$. The new momentum 
operators are given by $\tilde{q}_{x}=\sqrt{\frac{v_x}{v_y}}q_x$ and $\tilde{q}_{y}=\sqrt{\frac{v_y}{v_x}}q_y$
which satisfy the usual quantum mechanical commutation
relation $[\tilde{q}_x,\tilde{q}_y]=0$ and $[\tilde{x},\tilde{q}_x]=i$,
$[\tilde{y},\tilde{q}_y]=i$ provided $\tilde{x}=\sqrt{\frac{v_y}{v_x}}x$ and
$\tilde{y}=\sqrt{\frac{v_x}{v_y}}y$ . The corresponding energy dispersion is given by~\cite{goerbig2008tilted}
\begin{equation}\label{spectrum}
E({\tilde{q}_x}, {\tilde{q}_y})=\xi v_t \sqrt{\frac{v_x}{v_y}} {\tilde{q}_y} +\lambda v_F\mid \tilde{q}\mid\ .
\end{equation}

\begin{figure}[!thpb]
\centering
\includegraphics[width=0.92 \linewidth]{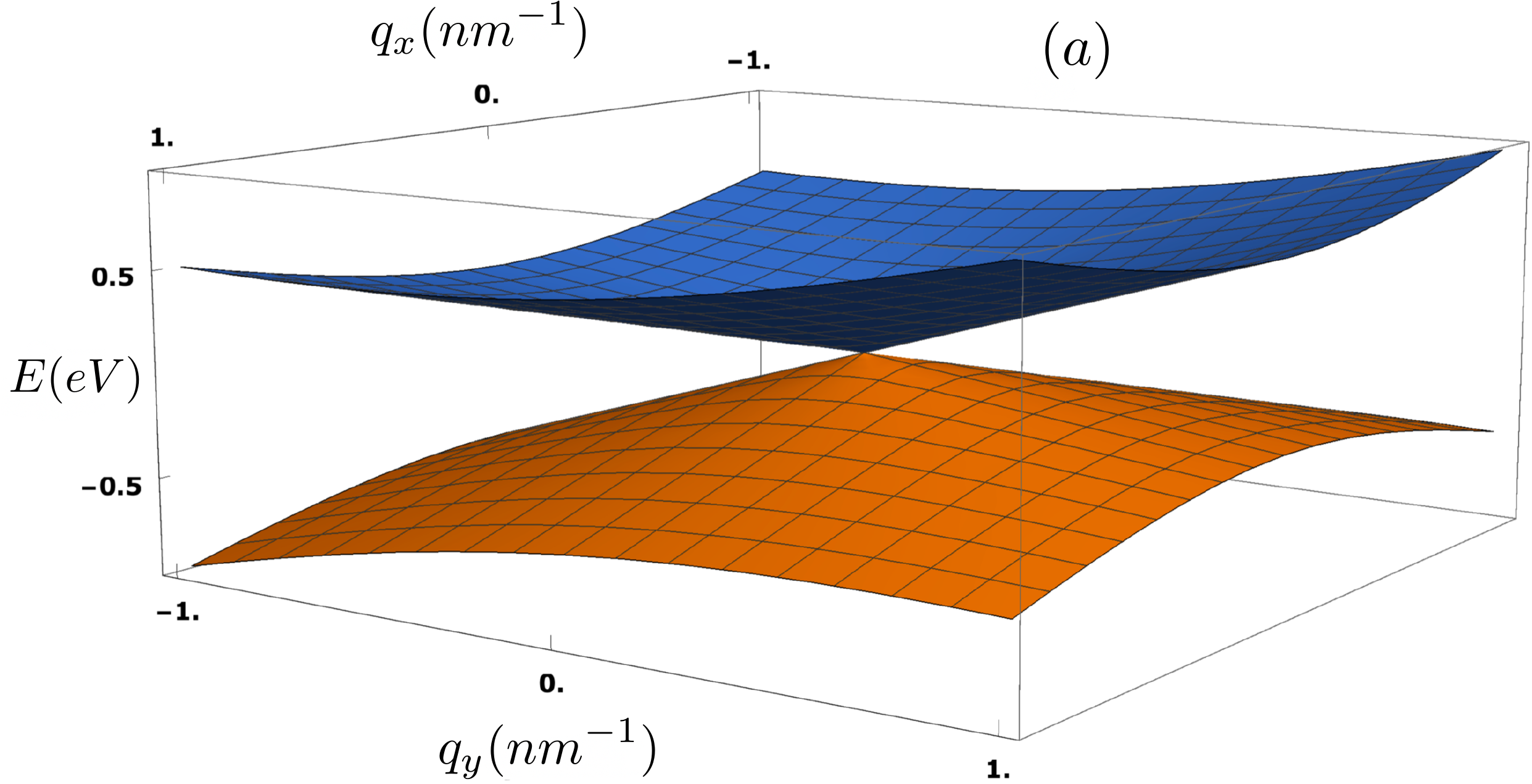}
\hfill
\includegraphics[width=0.9 \linewidth]{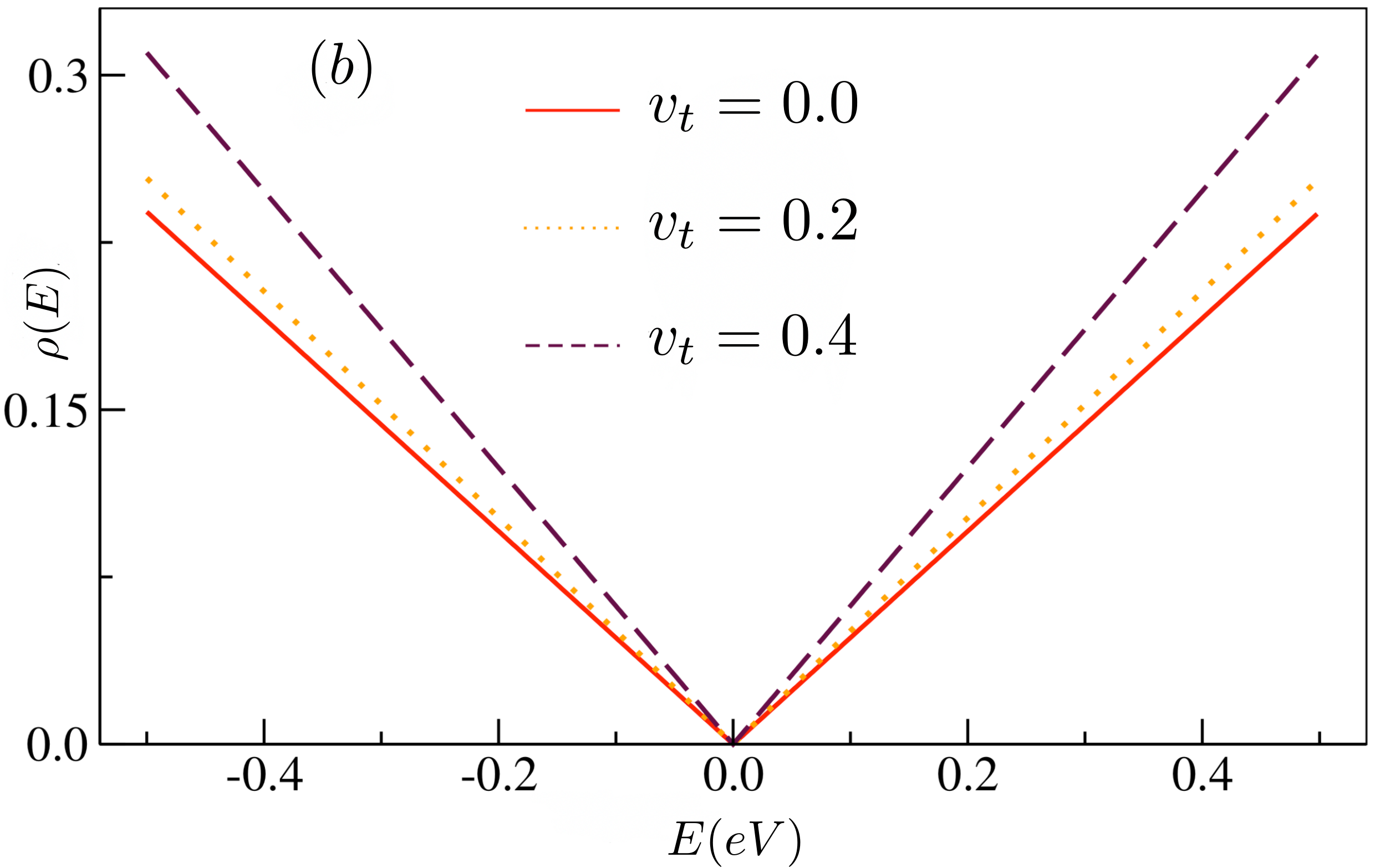}
\caption{(Color online) (a) A tilted, anisotropic Dirac cone of 8-Pmmn borophene, in the vicinity 
of Dirac point $\bf{k}_D$, is shown in the $k_x$-$k_y$ plane. (b) The behavior of DOS (in arbitrary unit)
is demonstrated near the Dirac point for different values of the tilt parameter $v_{t}$. }
\label{Fig2}
\end{figure}

%It is interesting to note that without tilting ($v_t=0$), the electronic band structure is still isotropic
%in a renormalized momentum space ($\tilde{q}_x,\tilde{q}_y$) with new Fermi velocity ($v_F$).
Here, $\lambda=\pm$ denotes band index. However, as we restrict ourself in the $n$-doped regime (conduction band),
therefore $\lambda$ is always positive in our analysis. The band structure near the Dirac point $\bf{k}=\bf{k_D}$
is shown in Fig.~\ref{Fig2}(a). The band dispersion around the other valley $\bf{k}=-\bf{k_D}$ has opposite chirality
\ie tilting lies in the opposite direction. It is important to note that the  tilting
breaks particle-hole symmetry in borophene~\cite{lopez2016electronic,zabolotskiy2016strain}.

Before proceeding further, we briefly examine how the tilting of the Dirac cones affects 
the Fermi energy and density of states (DOS). The Fermi energy ($E_F$) and DOS ($\rho(E)$)
in a material, associated with tilted and anisotropic Dirac cone, depend on the tilting
parameter $v_{t}$ in the following way
\bea
E_F(v_t)&=&E_F^{(0)} \Big(1-\frac{v_t^2}{v_F^2}\Big)^{\frac{3}{2}}\ ,\\
\label{EF}
\rho(E,v_t)&=&\rho^{(0)}(E) \Big(1-\frac{v_t^2}{v_F^2}\Big)^{-\frac{3}{2}}\ ,
\eea
where $E_F^{(0)}$ and $\rho^{(0)}(E)$ are the Fermi level and the DOS of a non-tilted isotropic Dirac material-\ie graphene,
respectively and $v_F=\sqrt{v_xv_y}$ as mentioned earlier. Note that, with the enhancement of the tilting parameter $v_{t}$, the Fermi level decreases monotonically. 
On the other hand, DOS gets enhanced with $v_{t}$ and the corresponding behavior of that, following 
the low energy spectrum (Eq.(\ref{spectrum})), is shown as a function 
of energy in Fig.~\ref{Fig2}(b). At the Dirac point, $\rho(E)$ vanishes and increases linearly
with energy similar to graphene~\cite{wallace1947band,bena2005quasiparticle}. Impact of these phenomena 
on the period of oscillation of RKKY exchange interaction is explained in the next section.

%----------------------------------------------------------------
%---------------------------------------------------------------------------------------------------------------------------- 
%-----------------------------------------------------------------------------------------------------------------------------
%---------------------------------------------------------------------
Now we briefly discuss the theoretical formalism for investigating the RKKY exchange interaction in $2$D 
electronic system. We consider two magnetic impurities (localized spins) at two different lattice
sites of the bulk 2D borophene sheet. The interaction term between the localized spins (${\bf{S}}_i$) and the conduction 
electron spins (${\bf{s}}_i$) is given by 
\bea
&&H_{int}=J ({\bf{S_1}}\cdot{\bf{s_1}}+{\bf{S_2}}\cdot{\bf{s_2}}) \ ,
\eea
where $J$ is the bare exchange coupling strength. Using second order perturbation theory, the exchange
interaction energy between the two localized spins can be written in the Heisenberg form as
\bea
&&E({\bf{r}})=J_{\alpha\beta}({\bf{r}}) {\bf{S_1}}\cdot{\bf{S_2}}\ , 
\eea
where $\alpha$, $\beta$ indicate the atomic indices and ${\bf{r}}$ is the distance between the two impurities.
Thus, the RKKY exchange interaction strength $J_{\alpha\beta}({\bf{r}})$ is given by 
\bea
&&J_{\alpha\beta}({\bf{r}})=\frac{J^2 \hbar^2}{4} \chi_{\alpha\beta}(0,{\bf{r}})\ .
\label{j_chi}
\eea
where $\chi_{\alpha\beta}(0,{\bf{r}})$ is the susceptibility.

For a spin-degenerate system, susceptibility can be written in terms of the unperturbed
retarded Green's functions as~\cite{sherafati2011analytical}
\begin{align}
&&\chi(r_1,r_2)=-\frac{2}{\pi}\int_{-\infty}^{E_F}\, dE~{\rm Im}[G_{\alpha\beta}^{(0)}(r_1,r_2,E)
G_{\beta\alpha}^{(0)}(r_2,r_1,E)]\ ,
\end{align}
To compute the susceptibility, which is directly proportional to RKKY exchange intercation, we need 
to evaluate the zeroth-order real space Green's functions. Since in the large distance, contribution to 
$\chi$  arises mainly from small momenta, one can extend the momentum cutoff to $\infty$ to obtain 
the real space Green's functions via the Fourier transform. We compute them in the linear band approximation and obtain
\bea
G_{\alpha\beta}^{(0)}({\bf{R}},0,E)&=&\frac{1}{\Omega_{BZ}}\int d^2{\bf{k}}~e^{i{\bf{k}}\cdot{\bf{R}}} G_{\alpha\beta}^{(0)}({\bf{k}},E)\non\\
&=&\frac{1}{\Omega_{BZ}}\int d^2{\bf{\tilde{q}}}~e^{i{\bf{\tilde{q}}}\cdot{\bf{R}}} [e^{i{\bf{K}}\cdot{\bf{R}}} G_{\alpha\beta}^{(0)}({\bf{\tilde{q}}}+{\bf{K}}, E)\non\\
&+&e^{i{\bf{K}}^{\prime}\cdot{\bf{R}}}G_{\alpha\beta}^{(0)}({\bf{\tilde{q}}}+{\bf{K}}^{\prime}, E)]\ ,
\label{GF1}
\eea
where the integration is performed over the entire Brillouin zone $\Omega_{BZ}$. Here, ${\bf\tilde{q}}=(\tilde{q}_{x}, \tilde{q}_{y})$
is small momentum in the vicinity of the Dirac points, where the linear Dirac spectrum is valid. As $d\tilde{q}_{x} d\tilde{q}_{y}\,=\,dq_{x}dq_{y}$, we can replace 
$\tilde{q}$ by $q$ without loss of generality. 
%which we have done throughout our analysis. 
We have also used the notation $(x,y)$ in place of $(\tilde{x},\tilde{y})$. The factor $\sqrt{\frac{v_x}{v_y}}$ in Eq.(\ref{spectrum}) has now been included in the tilting parameter $v_t$ for simplicity. The unperturbed momentum space Green's function is given by,
\begin{eqnarray}
G_{\alpha\beta}^{(0)}({\bf{k}},E)&=&(E+i\eta-H_D)^{-1}\non\\
&=&\frac{1}{D}
\begin{bmatrix}
E+i\eta-\xi v_tq_y & (q_x-iq_y) v_F\\
(q_x+iq_y) v_F & E+i\eta-\xi v_tq_y
\end{bmatrix},
\label{GF2}
\end{eqnarray}
with $D=(E+i\eta-\xi v_tq_y)^2-v_F^2 q^2 $, $v_F=\sqrt{v_xv_y}$. Here, $\xi=\pm 1$ indicates
the two Dirac points ${\bf{K}}$ and ${\bf{K^{\prime}}}$.
%The $q_x$ and $q_y$ used in Green's function are the normalized momentums. $\xi=\pm1$ indicate the Dirac points $K$ and $K^{'}$.

%----------------------------------------------------------------------
\section{Results}{\label{sec:III}}
%----------------------------------------------------------------------
In this section we present our analytical as well as numerical results for different locations of the two
magnetic impurities inside the bulk of the 2D borophene sheet. In general, it is a formidable task to
obtain the analytical expressions for the susceptibility when the two impurities are anywhere in the $x$-$y$ plane. 
%Obtaining analytical result for the susceptibility, when the impurities are randomly oriented, is difficult. 
However, we manage to obtain the analytical form of RKKY exchange interaction for the two special cases: 
(A) when the two impurities are located perpendicular to the tilt axis and (B) parallel to the tilt axis.  Here, we clarify our notations used for our analysis: $\chi_{11}$ denotes the susceptibility
for the impurities being on same type of atoms and $\chi_{12}$ for the different types of atoms (one on ridge atom and the other on inner atom (see Fig.~\ref{model})). It is worthwhile to mention that 
when the impurities are on different atoms, they cannot be exactly along $x$ or $y$ axis. However, such small off-axis deviation along the tilt axis ($y$ axis) or perpendicular to it ($x$ axis), 
does not change our results qualitatively, in the continum limit. 
%affect much as this deviation effect decreases with distance between the impurities}. 
Note that, RKKY exchange interaction strength is directly proportional to the susceptibility (see Eq.(\ref{j_chi})). Hence, we express the susceptibility as a measure of RKKY interaction strength 
in units of $J^{2}\hbar^{2}/4$. Also, we normalize the distance between the two magnetic impurities by corresponding lattice parameters \ie $R_x/a\to R_x$ and $R_y/b\to R_y$.

%From Eq.(\ref{j_chi}), we note that RKKY exchange interaction strength is directly proportional to the susceptibility. Now on we present susceptibility as a measure of RKKY interaction. We normalize the distance by corresponding lattice parameter \ie $R_x/a\to R_x$ and $R_y/b\to R_y$.
%----------------------------------------------------------------------
\subsection{Impurities are located perpendicular to the tilt axis} 
%----------------------------------------------------------------------
\vspace{0.1cm}
When the two magnetic impurities reside perpendicular to the tilt axis (on the same atoms) of bulk borophene, 
the integral in Eq.(\ref{GF1}) can be computed analytically for $R_y=0$. This configuration has
been indicated by golden arrows in Fig.~\ref{model}. In this case, the zeroth-order real space Green's function reads as
%\ie when the impurities are along the x axis as shown by golden arrows in Fig.~\ref{model}. We now present the result when the impurities are on the same atom.
\bea
&&G^{(0)}_{11}(R_x,0,E)\non\\
&=&\frac{1}{\Omega_{BZ}}\int dq_x\,dq_y \frac{(E+i\eta-v_tq_y \xi) e^{iq_xR_x}}{(E+i\eta-v_tq_y\xi)^2-v_F^2(q_x^2+q_y^2)}\non\\
&=&\frac{-2\pi E}{v_F}\frac{(v_F^2-2v_t^2)}{(v_F^2-v_t^2)^{\frac{3}{2}}} K_0(-i\tilde{\alpha}) (e^{i{\bf{{K}}}\cdot{\bf{{R}}}}+e^{i{\bf{K^{\prime}}}\cdot{\bf{R}}})\ , 
\eea
%\bea
%G_{11}^I(R,0,E)&=&\int dq_x\,dq_y \frac{(E+i\eta) e^{i(q_xR_x+q_yR_y)}}{(E+i\eta-v_tq_y\xi)^2-v_F^2(q_x^2+q_y^2)}\non\\
%&=&\frac{E\pi^2}{v_F}(iJ_0(\tilde{\alpha})+Y_0(\tilde{\alpha})) (e^{i\vec{K}.\vec{R}}+e^{i\vec{K^{'}}.\vec{R}})
%\eea
with $\tilde{\alpha}=ER_{x}/\sqrt{v_F^2-v_t^2}$. Here, $K_0(x)$ are the modified Bessel function of first kind. Expressing the modified Bessel function in terms of Bessel and Neumann function, we obtain
\bea
&&{\rm{Im[G^{(0)}_{11}}}(R_x,0,E)G^{(0)}_{11}(0,R_x,E)]=\frac{4 \pi^2E^2}{\Omega_{BZ}^2}\frac{v_F^2}{(v_F^2-v_t^2)^3}\non \\
&& \times {\rm {Im}}[(K_0(-i\tilde{\alpha}))]^2(2+2\cos\{(K_x-K_x^{'})R_x\})\ .
\eea

Henceforth, following the works of Saremi~\cite{saremi2007rkky} and Sherafati \etal~\cite{sherafati2011rkky,sherafati2011analytical}, we separate the integration limit : $\int_{-\infty}^{E_F}=\int_{-\infty}^{0}+\int_{0}^{E_F}$. The first integral indicates the valence electrons (undoped case) and the second one is for the conduction electrons. While the latter integral involves Meijer G-function, former one does not converge. Following the standard procedure~\cite{saremi2007rkky,sherafati2011analytical}, the integrand can be multiplied by a cutoff function $f(\tilde{\alpha},\tilde{\alpha}_0)$=$\exp{(-\tilde{\alpha}/\tilde{\alpha}_0)}$. Then one can perform the integral and takes limit $\tilde{\alpha}_0\to \infty$ at the end so that $f(\tilde{\alpha},\tilde{\alpha}_0)\,\to\,1$.  Thus, we arrive at the following form of 
the susceptibility:
\bea
\chi_{11}&=&\frac{1}{\pi R_x^3}\frac{v_F^2}{(v_F^2-v_t^2)^{\frac{3}{2}}}[1+\cos\{(K_x-K_x^{\prime})R_x\}]\non\\
&&\times \Bigg[\frac{1}{16}-\frac{k_F^{'} R_x}{2\sqrt{\pi}}M(k_F^{'} R_x)\Bigg]\ ,
\label{chi11}
\eea
where, $k_F^{'}=E_F/\sqrt{v_F^2-v_t^2}$ and $M(k_F^{'} R_x)$=$\MeijerG*{2}{0}{1}{3}{\frac{3}{2}}{1,1,1,-\frac{1}{2}}{{k_F^{'}}^2 R_x^2}$ is the Meijer G-function. We have considered the first Brillouin zone 
area as $\Omega=4\pi^2/ab$. Tilted anisotropic Dirac points are at $\vec{k}_D=(0,k_D)$ and $-\vec{k}_D$, $k_D=0.29\times\frac{2\pi}{b}$~\cite{zabolotskiy2016strain}. It is interesting to note that in case of borophene, the oscillatory factor $[1+\cos\{(K_{x}-K_{x}^{\prime})R_x\}]=2$. Hence, the interference terms between the two Dirac points do not contribute to the RKKY exchange interaction which is evident 
from Eq.(\ref{chi11}). 

\begin{figure}[!thpb]
\centering
\includegraphics[width=0.5\textwidth]{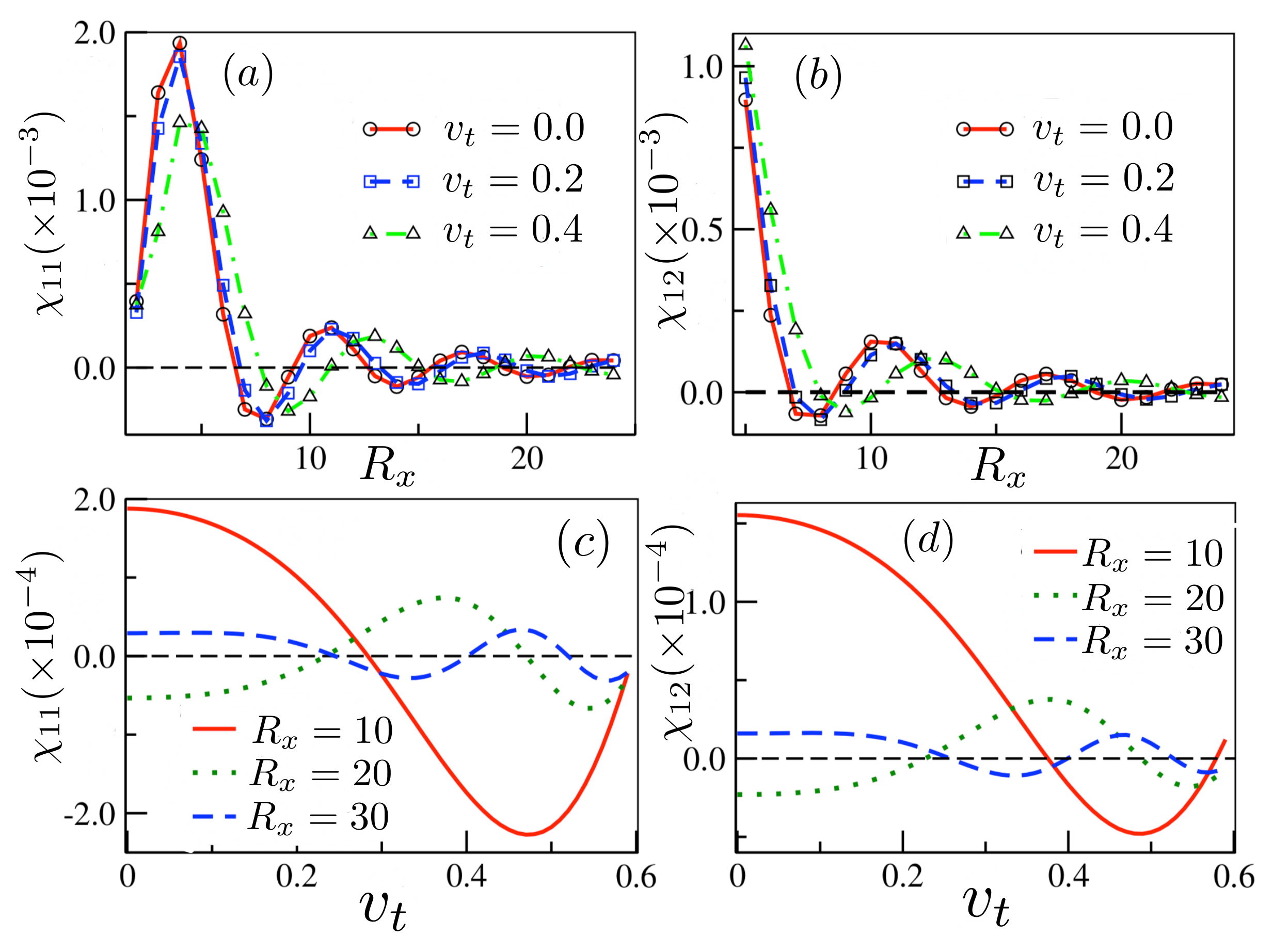}
\caption{(Color online) The behavior of the suscepibilities $\chi_{11}$ and $\chi_{12}$ is shown as a function of $R_x$ in panels (a) and (b), and as a function of $v_t$ in panels (c) and (d) respectively. 
We choose the other parameters as $v_F=1.0$, $R_y=0$, $E_F^{(0)}=0.5$. Black dashed line indicates the null susceptibility.}
\label{Fig3}
\end{figure}

In Fig.~\ref{Fig3}, we demonstrate the behavior of susceptibility $\chi_{11}$ (when impurities are on the same atom) as a function of the distance $R_x$ between the two magnetic impurities 
and tilt parameter $v_t$. The change in the periodicity of $\chi_{11}$, with the enhancement of tilting parameter $v_t$, is evident from Fig.~\ref{Fig3}(a).
%has been presented as a function of $R_x$ and $v_t$ in Fig.~\ref{Fig3}. 
%We can see the change in the periodicity of $\chi_{11}$, with the increase of tilting parameter $v_t$, in Fig.~\ref{Fig3}(a). 
This feature can be understood from Eq.(\ref{chi11}) in which the Meijer G-function is the entire source of the oscillation. As the period of oscillation is inversly proportional to the argument $k_F^{\prime} R_x$, 
it scales with $v_t$ as $\tau(v_t)=\tau(0)/(1-x^2)$, where $x=v_t/v_F$ and $\tau(0)$ is the period of non-tilted and isotropic Dirac cone \ie graphene. As the tilting parameter increases, period of oscillation
increases monotonically. In the $v_t\to 0$ limit, $\tau(v_t)$ comes back to the untilted period $\tau(v_t=0)$ as expected. On the other hand, for $v_t\to v_F$, the period diverges indicating flatness of exchange interaction. Hence, the internal band structure itself influences the RKKY exchange (Friedel oscillation) due to tilting and may be a way to probe the degree of tilting in anisotropic 2D Dirac materials. 
This is also one of the main results of our paper.
%This feature is interesting as the internal band structure influences the Friedel oscillation due to tilting.

%------------------------------------------------------------------------------------------
%------------------------------------------------------------------------------------------
%----------------------------------------------------------------------------------------
%----------------------------------------------------------------------------------------

We also observe that there is a reduction in the amplitude of $\chi_{11}$ as $v_t$ increases (see Fig.~\ref{Fig3}(a)). The reason can be attributed to the presence of vanishingly small 
density of states near the Dirac point and lowering of Fermi level with the rise of tilting parameter $v_{t}$. 
%which can be attributed to the lesser density of state near the Dirac point and lowering of Fermi level with the rise of tilting parameter. 
Furthermore, we explore the long distance and very short distance limit of RKKY exchange interaction which can be figured out from the asympototic behavior of Meijer G-function. 
Using the standard tables~\cite{meijer1946g,fields1972asymptotic} and following Ref.~[\onlinecite{sherafati2011analytical}], we obtain
\bea
\displaystyle{\lim_{y\to 0}}M(y)&=&\frac{4y^2[1-3\gamma-3 \ln(y/2)]}{9\sqrt{\pi}}\ , \\
\displaystyle{\lim_{y\to \infty}}M(y)&=&\frac{[2\cos(2y)+8y\sin(2y)-\pi]}{8\sqrt{\pi}y}\ ,
\label{limitM}
\eea
where $\gamma\approx 0.577$ is the Euler-Mascheroni constant. Therefore, we obtain the following form of the RKKY exchange interation in the long distance limit ($k_F^{'}R_x\gg1$) as
\bea 
\displaystyle{\lim_{k_F^{'}R_x\to \infty}}\chi_{11}(R_x)&=&\frac{\chi_{11}^{L}}{R_x^3}[\pi-\cos(2 k_F^{'}R_x)\non\\&&-4k_F^{'}R_x\sin(2 k_F^{'}R_x)]\ ,
\eea
where $\chi_{11}^{L}=v_F^2/8\pi^2(v_F^2-v_t^2)^{3/2}$. On the other hand, our analytical results for the short distance limit ($k_F^{\prime}R_x\ll1$) reads
%In the very small distance limit ($k_F^{'}R_x<<1$), we find
\bea
\displaystyle{\lim_{k_F^{'}R_x\to 0}}\chi_{11}(R_x)&=&\frac{\chi_{11}^{S}}{R_x^3}\Bigg[1-\frac{32 (k_F^{'}R_x)^3}{9\pi}\non\\
&&(1-3\gamma-3\ln(k_F^{'}R_x/2))\Bigg]\ .
\eea
where $\chi_{11}^{S}=v_F^2/16\pi(v_F^2-v_t^2)^\frac{3}{2}$.

Here we present the results for the magnetic impurities located on different atoms. One can obtain the Green's function extending the momentum cutoff to $\infty$ as
\bea
&&G^{(0)}_{12}(R_{x},0,E)\non\\
&=&\frac{1}{\Omega_{BZ}}\int dq_x\,dq_y \frac{(q_x-iq_y) e^{iq_xR_x}}{(E+i\eta-v_tq_y\xi)^2-v_F^2(q_x^2+q_y^2)}\non\\
&=&-\frac{2\pi E}{\Omega_{BZ} (v_F^2-2v_t^2)}K_1(-i\tilde{\alpha}) (e^{i{\bf{K}}\cdot{\bf{R}}}+e^{i{\bf{K^{\prime}}}\cdot{\bf{R}}})\non\\
&&-\frac{2i\xi v_t \pi E}{\Omega_{BZ} (v_F^2-2v_t^2)^{\frac{3}{2}}}K_0(-i\tilde{\alpha}) (e^{i{\bf{K}}\cdot{\bf{R}}}-e^{i{\bf{K^{\prime}}}\cdot{\bf{R}}})\,
\eea
Similar to $\chi_{11}$, the integrand is multiplied by a cutoff function to evaluate the energy integral of the valence electrons. Finally we obtain
\bea
\chi_{12}&=-&\frac{1}{2\pi R_x^3 \sqrt{v_F^2-v_t^2}}\Bigg(-\frac{3}{16}-\frac{k_F^{'} R_x}{2\sqrt{\pi}}\tilde{M}(k_F^{'} R_x)\Bigg)\non\\
&&\times [1+\cos\{(K_{x}-K_{x}^{\prime})R_x\}]\non\\
&-&\frac{v_t^2}{2\pi R_x^3 (v_F^2-v_t^2)^{\frac{3}{2}}}\Bigg(\frac{1}{16}-\frac{k_F^{'} R_x}{2\sqrt{\pi}}M(k_F^{'} R_x)\Bigg)\non\\
&&\times [1-\cos\{K_{x}-K_{x}^{\prime})R_x\}]\ ,
\label{chi12}
\eea
where $\tilde{M}(k_F^{'} R_x)=\MeijerG*{2}{1}{2}{4}{\frac{1}{2},\frac{3}{2}}{1,2,0,-\frac{1}{2}}{{k_F^{'}}^2 R_x^2}$.
In Eq.(\ref{chi12}), the second term vanishes for borophene ($K_{x}=K_{x}^{\prime}=0$), while in the first term, again the interference between the anisotropic Dirac cones does not contribute 
to the RKKY oscillations. Note that, both $\chi_{11}$ and $\chi_{12}$ recover the similar form of graphene~\cite{sherafati2011analytical,sherafati2011rkky} in the limit : $v_t=0,\, v_x=v_y=v_F$.

Behavior of $\chi_{12}$, as a function of $R_x$ and $v_{t}$, is depicted in Fig.~\ref{Fig3}(b).
Here also the period of oscillation increases with the increment of $v_t$, similar to $\chi_{11}$. 
This feature can be understood from Eq.(\ref{chi12}) where the argument of Meijer G-function
is proportional to $v_t$. We also explore the long distance ($k_F^{'}R_x\gg1$) 
and short distance ($k_F^{\prime}R_x\ll1$) limit of $\chi_{12}$ employing asymptotic behavior
of Meijer G-function. We find,
\bea
\displaystyle{\lim_{y\to 0}}\tilde{M}(y)&=&\frac{2y^2}{3\sqrt{\pi}}\ , \\
\displaystyle{\lim_{y\to \infty}}\tilde{M}(y)&=&\frac{[3\pi-10\cos(2y)-8y\sin(2y)]}{8\sqrt{\pi}y}\ .
\label{limitM}
\eea
Employing the above limits, we obtain the following analytical forms of RKKY exchange interaction as
\bea
\displaystyle{\lim_{k_F^{'}R_x\to \infty}}\chi_{12}(R_x)&=&\frac{\chi_{12}^{L}}{R_x^3}[3\pi-5\cos(2 k_F^{'}R_x)\non\\
&&-4k_F^{'}R_x\sin(2 k_F^{'}R_x)]\ , \\
\displaystyle{\lim_{k_F^{'}R_x\to 0}}\chi_{12}(R_x)&=&\frac{\chi_{12}^{S}}{R_x^3}\Bigg[1+\frac{16(k_F^{'}R_x)^3}{9\pi}\Bigg]\ ,
\eea
with $\chi_{12}^{L}=1/8\pi^2 \sqrt{v_F^2-v_t^2}$ and $\chi_{12}^{S}=3/16\pi \sqrt{v_F^2-v_t^2}$. 

The oscillatory behavior of $\chi_{11}$ and $\chi_{12}$ as a function of $v_t$ is clearly visible
from Figs.~\ref{Fig3}[(c)-(d)]. They exhibit similar features for $\chi_{11}$ and $\chi_{12}$ with the 
variation of $v_t$ as in both cases the period of oscillation depends on the tilting parameter.
%Features of $\chi_{11}$ and $\chi_{12}$ as a function of $v_t$ are almost similar. 
%{\textcolor{red}{Also note that, the period of oscillation increases with the separation between the two impurities as the oscillation frequency is proportional to $R_x$.}} 
It is evident that due to $R_x^{-3}$ dependency (see Eq.(\ref{chi11}) and Eq.(\ref{chi12})),
amplitude of both $\chi_{11}$ and $\chi_{12}$ become vanishingly small as one gradually enhances 
the distance between the two magnetic impurities.

\begin{figure}[!thpb]
\centering
\includegraphics[width=0.5\textwidth]{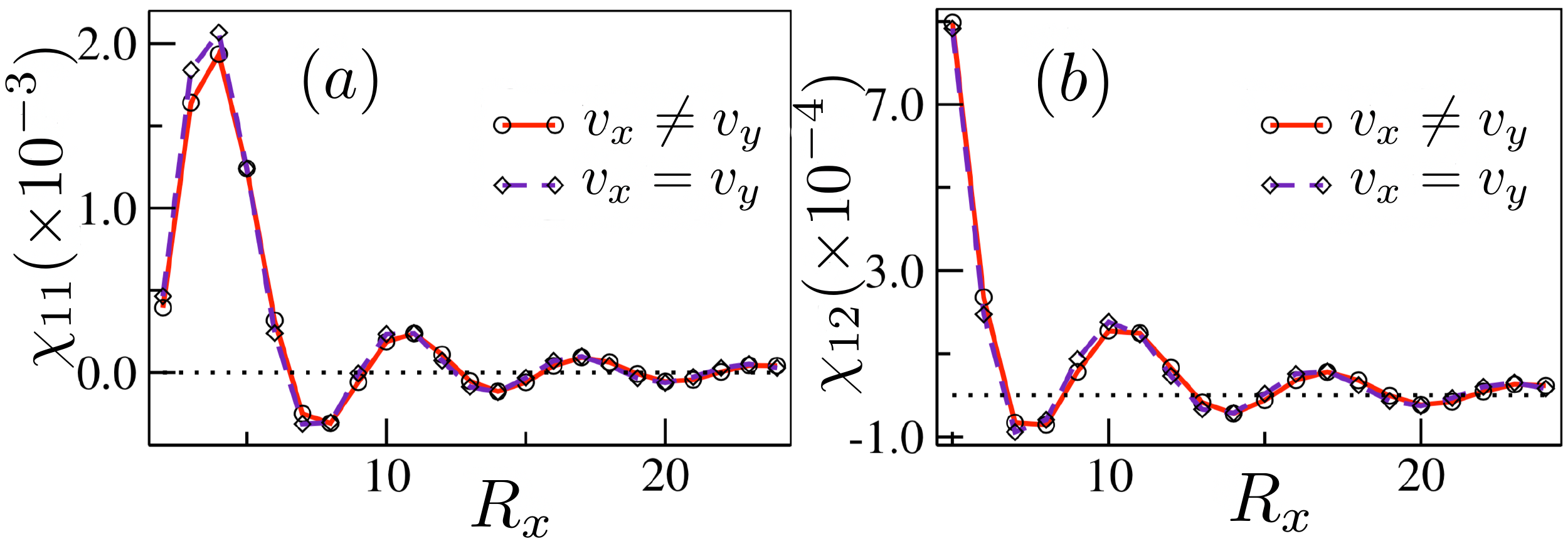}
\caption{(Color online) The behavior of the suscepibilities $\chi_{11}$
and $\chi_{12}$ are illustrated as a function of $R_x$ with $v_t=0$ for
both anisotropic Fermi velocities and isotropic counterpart in panel (a) 
and (b) respectively. The value of the other parameters are chosen as $R_y=0$, $E_F^{(0)}=0.5$.}
\label{Fig4}
\end{figure}

It is important to note that, even without the tilting parameter, the Fermi surface is anisotropic
due to the anisotropy in Fermi velocities ($v_x\neq v_y$). Effect of this anisotropy, in the absence of tilting 
($v_{t}=0$), on RKKY exchange interaction is presented in Fig.~\ref{Fig4}. It is evident that the
effect of such simple anisotropic Fermi surface on susceptibility ($\chi_{11}$ and $\chi_{12}$) is 
negligibly small. This may be a possible way to distinguish between these two kind of anisotropies
($v_{t}=0$ and $v_{t}\neq 0$) by measuring the RKKY interaction. 

%-----------------------------------------------------------------------------------------------------------------------
%---------------------------------------------------------------------------------------------------------------------

% {\textcolor{blue}{It is important to mention that in absence of any tilting ($v_t=0$),
% RKKY exchange interaction is not affected by the anisotropy in the Fermi velocities 
% ($v_x\neq v_y$). This can be understood as the following: the Hamiltonian in Eq.(\ref{Ham}) can be written as}
% \begin{equation}
% H_D=\xi\{v_F( \sigma_x \tilde{q}_x +\sigma_y \tilde{q}_y)\}\non
% \end{equation}
% 
% {\textcolor{blue}{with $v_F=\sqrt{v_x v_y}$. Now the anistropies are in the momenta
% $\tilde{q_x}$ and $\tilde{q_y}$ where $\tilde{q}_x=q_x\sqrt{\frac{v_x}{v_y}}$ and 
% $\tilde{q}_y=q_y\sqrt{\frac{v_y}{v_x}}$. To obtain  zeroth order real space Green's 
% function, as mentioned earlier, the Green's function in momentum space has to be integrated 
% over both $q_x$ and $q_y$.  The Jacobian is $1$ in this case \ie $d\tilde{q}_x d\tilde{q}_y=dq_x dq_y$.
% Hence after the integrations, this anisotropy in momentum space do not play any role
% in determining the susceptibility. In other words, there is no affect of the anisotropy
% in Fermi surface on the period of oscillation and thus the increament of period of oscillation,
% when the impurities are perpendicular to the tilt axis, is solely due the to tilt parameter
% which one of the main focus of our paper.}

%----------------------------------------------------------------------
\subsection{Impurities are aligned parallel to the tilt axis}
%----------------------------------------------------------------------
%\vspace{0.1cm}
Here, we present the analytical form of the susceptibility in the limit $R_x\to 0$ \ie
when the two magnetic impurities are situated along the tilt axis. This has been illustrated by
golden and dark green arrows in Fig.~\ref{model}. We proceed in the similar way as in the previous 
subsection and obtain
\bea
\chi_{11}&=&-\frac{1}{\pi R_y^3 v_F}  \int_{-\infty}^{z_F} dz z^2\Bigg[-{\cos\Bigg(K_y R_y-\frac{v_t}{v_F}z\Bigg)}^2 J_0(z)Y_0(z)\non\\
&&+\frac{v_t^2}{v_F^2}{\sin\Bigg(K_y R_y-\frac{v_t}{v_F}z\Bigg)}^2 J_1(z)Y_1(z)\Bigg]  \ ,
\label{chi11_Rx0}           
\eea
where, $z=Ev_FR_y/(v_F^2-v_t^2)$ and $z_F=E_Fv_FR_y/(v_F^2-v_t^2)$.
\bea
\chi_{12}&=&-\frac{1}{\pi R_y^3 v_F}  \int_{-\infty}^{z_F} dz z^2\Bigg[{\cos\Bigg(K_y R_y-\frac{v_t}{v_F}z\Bigg)}^2 J_1(z)Y_1(z)\non\\
&&-\frac{v_t^2}{v_F^2}{\sin\Bigg(K_y R_y-\frac{v_t}{v_F}z\Bigg)}^2 J_0(z)Y_0(z)\Bigg] \ .
\label{chi12_Rx0}        
\eea

\begin{figure}[!thpb]
\centering
\includegraphics[width=0.5\textwidth]{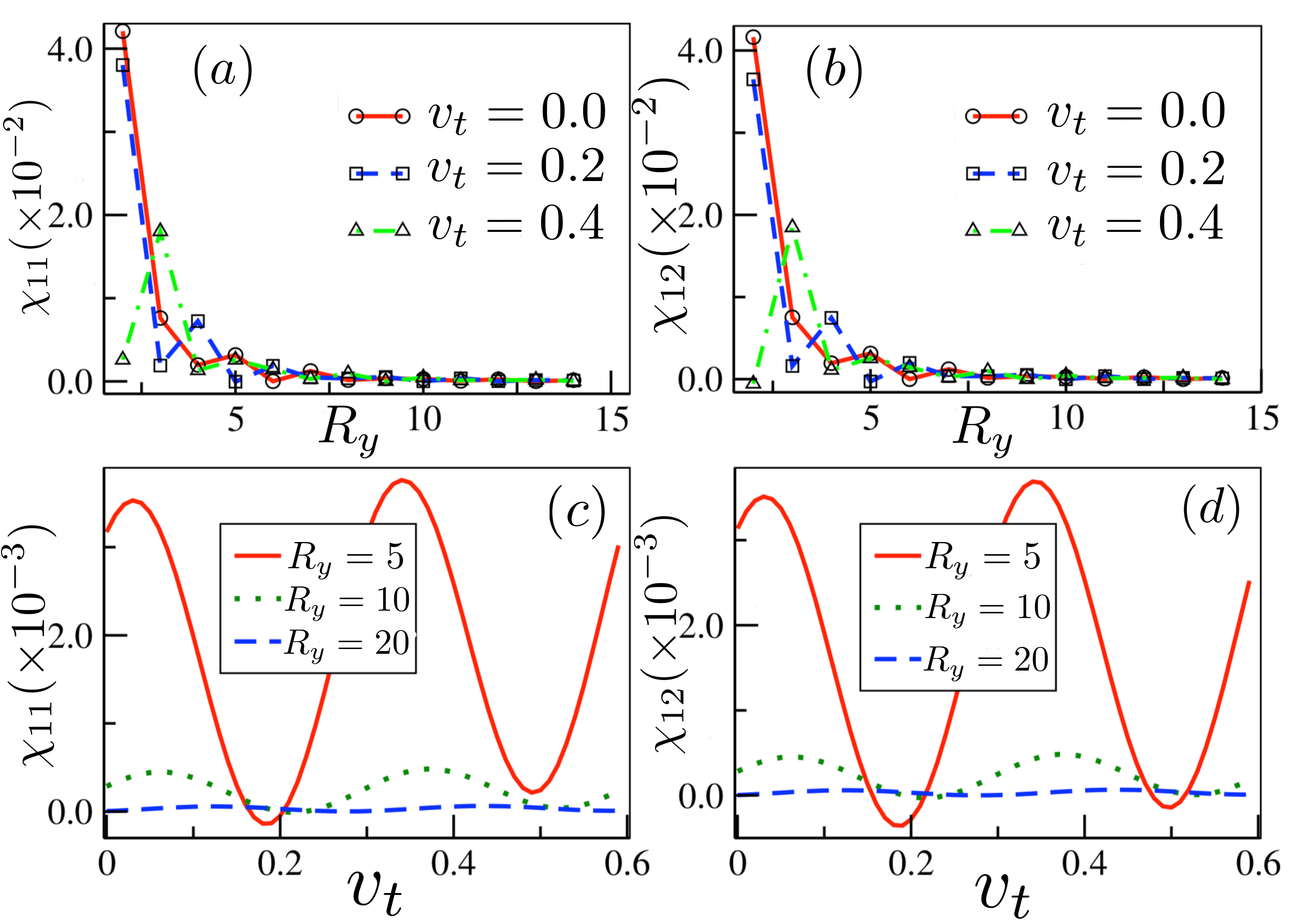}
\caption{(Color online) The behavior of the suscepibilities $\chi_{11}$ and $\chi_{12}$
is illustrated as a function of $R_y$ in panels (a) and (b), and as a function of $v_t$
in panels (c) and (d) respectively. The value of the other parameters are chosen as
$v_F=1.0$, $R_x=0$, $E_F^{(0)}=0.5$.}
\label{Fig5}
\end{figure}

The explicit analytical form of both $\chi_{11}$ and $\chi_{12}$, after performing the
integration over energy, are very messy and not possible to write in a compact form. Rather, we present our results 
for them in Fig.~\ref{Fig5} both as a function of $R_y$ and $v_t$. We note that Fig.~\ref{Fig5}(a)
is almost identical to Fig.~\ref{Fig5}(b) and Fig.~\ref{Fig5}(c) is to Fig.~\ref{Fig5}(d). This feature can be attributed to the fact that the Hamiltonian is written in two atomic basis (two sublattice indices) of borophene (tilted Graphene), which do not couple directly to the magnetic moments. Such feature is in complete contrast to the case of topological insulator~\cite{liu2009magnetic,zhu2011electrically} in which the actual spin degree of freedom of the surface states directly couples to the impurity spin moment.

%-----------------------------------------------------------------------------------------------------------------------
%-----------------------------------------------------------------------------------------------------------------------
%---------------------------------------------------------------------------------------------------------------------
%---------------------------------------------------------------------------------------------------------------------

We observe that the period of oscillation is almost constant with $v_t$, as can be seen from Fig~\ref{Fig5}(c).
In this particular limit of $R_x\to 0$, the interference among the Dirac fermions from different Dirac points in
the Brillouin zone plays a crucial rule in determining the oscillatory nature of RKKY exchange interaction.
From Eqs.[(\ref{chi11_Rx0})-(\ref{chi12_Rx0})], one can note that the period of oscillation depends both
on $K_y$ and $v_t$. As the tilting angles (velocity) lie in opposite directions for the two Dirac 
points $\bf{K}$ and $\bf{K}^{'}$, the effect of tilting on RKKY exchange interaction, nullifies each
other due to destructive interference. Hence, the period of oscillation does not vary with the enhancement of $v_t$.
%{\textcolor{blue}{As the tilting lies in opposite direction for the two Dirac points $\bf{K}$ and $\bf{K}^{'}$,
%the tilting effect on RKKY exchange interaction, impurities being along the tilt direction, cancels each other.} Hence, the period of oscillation does not vary with the enhancement of $v_t$. 
The latter behavior indicates that the interference between the anisotropic Dirac cones dominates over the effect of Fermi level in determining the overall period of oscillation. 
Nevertheless, the amplitude of RKKY oscillation decreases with the degree of tilting $v_{t}$ due to the lowering of the Fermi level
(see Eq.(\ref{EF})). Although DOS increases with the tilting parameter, however near the Dirac points DOS
is negligbly small and hence it contributes negligibly in determining the amplitude
of oscillation, especially in the long distance limit where only small momenta are important.
The presence of interfence among the Dirac points enhances the oscillation frequency of
both $\chi_{11}$ and $\chi_{12}$ compared to the previous case (see Figs.~\ref{Fig3}[(a)-(b)]).
Also, the oscillatory behavior of RKKY exchange with enhanced frequency as a function of $v_t$, is clearly visible 
from Fig.~\ref{Fig5}(c) and Fig.~\ref{Fig5}(d).

%----------------------------------------------------------------------------------------
%------------------------------------------------------------------------------------------
\begin{figure}
\centering
\includegraphics[width=0.4\textwidth]{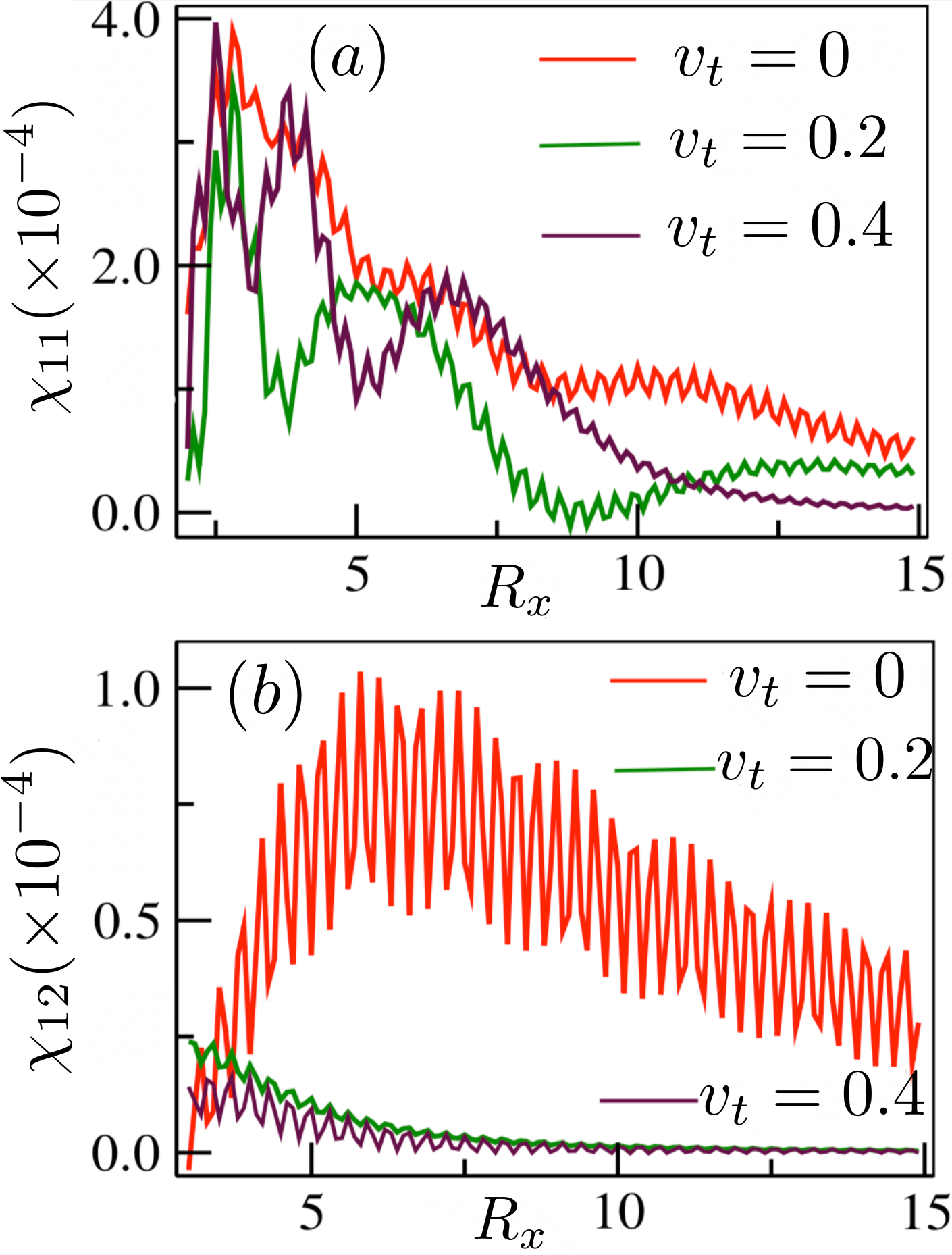}
\caption{(Color online) The behavior of the susceptibilities $\chi_{11}$ and $\chi_{12}$ are depicted as a function of $R_x$ in panels (a) and (b) respectively. 
The value of the other parameters are chosen as $v_F=1.0$, $E_F^{(0)}=0.5$, $R_y=5$ in panel (a) and $R_y=2$ in panel (b).}
\label{Fig6}
\end{figure}
%----------------------------------------------------------------------------------------
%----------------------------------------------------------------------------------------

%----------------------------------------------------------------------
\subsection{Impurities are located in the $x$-$y$ plane} 
%----------------------------------------------------------------------
In this subsection, we present our numerical results when the two magnetic impurities are placed at arbitrary positions in the $x$-$y$ plane of the bulk 2D borophene sheet \ie $R_x\neq 0$ and $R_y\neq 0$. 
In this case, it is not possible to obtain any kind of analytical form of the susceptibility. 
Hence, we compute the real space Green's function by numerically integrating over momenta and 
energy degrees of freedom. The corresponding results for $\chi_{11}$ and $\chi_{12}$ are 
illustrated as a function of $R_x$ for different values of the tilting parameter $v_{t}$ 
with $R_y=5$ in Fig.~\ref{Fig6}(a) and
$R_y=2$ in Fig.~\ref{Fig6}(b) respectively.
%Green's function is numerically integrated over momenta and energy degrees of freedom and 
%the results are shown as a function of $R_x$ for different tilting parameter in Fig.~\ref{Fig6} with $R_y=5$ in (a) and $R_y=2$ in (b). 
It is apparent from Fig.~\ref{Fig6}(a) (when two magnetic impurities are placed on same
atoms in the $x$-$y$ plane) that the oscillations appear very rapidly as a function of $R_x$
and manifest weak dependence on the values of $v_t$. These high frequency rapid oscillations emerge due to 
the interference between the two Dirac cones. This has been reported earlier for other 
2D Dirac systems~\cite{black2010rkky,sherafati2011rkky,islam2018probing}. The competition between the 
enhancement of DOS and reduction of Fermi energy with tilting parameter (see Eq.(\ref{EF})), 
for arbitrarily placed impurities, results in the non-monotonic variation of amplitude
of oscillation as a function of $v_{t}$ in the short and intermediate distance scale. This feature can be seen from 
Fig.~\ref{Fig6}(a). However for large distance between the two impurities, only small momenta
contribute to the RKKY exchange interaction and hence lowering of the Fermi level controls
the amplitude, resulting in monotonic decrease of $\chi_{11}$ with tilting.

The behavior of the spin density oscillations, for two magnetic impurities placed on different atomic sites
in the $x$-$y$ plane, is shown in Fig.~\ref{Fig6}(b) choosing $R_y=2$. There is one notable
difference between Fig~\ref{Fig6}(a) and Fig.~\ref{Fig6}(b) is that in Fig~\ref{Fig6}(a), the
oscillatory behavior associated with an envelope of the amplitude is prominent while in Fig.~\ref{Fig6}(b) 
the period is larger. As the period of oscillation is inversely proportional to the distance, 
for very small distance, envelope of amplitude of the RKKY exchange interaction decays
(period becomes large) while the oscillations are prominent for larger distances. 
Moreover, for the small separation of the impurities, $\chi_{12}$ exhibits non monotonic
behavior with $v_t$. On the other hand, for large distance ($R_x>5$) 
$\chi_{12}$ decreases monotonically with the tilting parameter $v_{t}$ which can again
be attributed to the effect of tilting on the Fermi level and DOS.

%----------------------------------------------------------------------
\section{Summary and Conclusions}{\label{sec:IV}}
%----------------------------------------------------------------------
%In the earlier works of RKKY interaction in graphene, continuum model~\cite{saremi2007rkky,brey2007diluted} has mostly been used with linearly dispersive band structure having two Dirac cones in Brillouin zone. Also in a finite size lattice of Graphene, exact diagonalization method has been applied~\cite{black2010rkky}. Sherafati \etal provided analytical expression~\cite{sherafati2011rkky,sherafati2011analytical} of RKKY interaction in graphene in terms of Meijer G-function. However, there are disagreements in the results because of various approximations used in those works such as using sharp cut off function to remove divergency, finte size effcet etc.

To summarize, in this article, we have explored the effect of the tilted and anisotropic Dirac
cones on the RKKY exchange interaction in the bulk of a $8$-Pmmn borophene. We observe that the 
tilting of the Dirac cones, for specific orientation of the impurities, manifests itself with the
significant reduction in the RKKY exchange interaction oscillation frequency. This feature can be
an indirect signature of the degree of tilting present in tilted and anisotropic Dirac cone.
We present our analytical expressions for the susceptibility, in terms of 
Meijer G-function, which is directly proportional to the exchange interaction strength.
We consider two special cases: when two magnetic impurities are located perpendicular
to the tilt axis ($x$ axis) and along the tilt axis ($y$ axis). In the former case, interference between the Dirac cones does not
contribute to the Friedel oscillations and the period of oscillation increases with
tilting parameter. In contrast, for the impurities being along the tilt axis, interference
among the Dirac cones plays the dominant role in determining the period of oscillation
while the tilting parameter exhibits negligible contribution. Moreover, due to opposite orientation of tilting of the Dirac 
cones at the inequivalent Dirac points, the effect of tilting originating from each Dirac point on RKKY exchange interaction 
nullifies each other when the impurities reside along the tilt axis. However, the amplitude of oscillation decreases with 
the tilting parameter beacuse of lowering of the Fermi level. We also separate out the effect of simple anisotropy ($v_{x}\neq v_{y}$)
on RKKY exchange interaction in absence of tilting ($v_{t}=0$) and show that such anisotropy in Fermi surface exhibits negligible effect 
on the response function in case of borophene. For arbitrarily placed magnetic impurities in the $x$-$y$ plane (neither along nor perpendicular to the tilt axis), 
we evaluate our results numerically and show rapid oscillations (beating pattern) in susceptibility due to the interference between the two Dirac cones 
and subdominant effects arising from the tilt parameter.

%numerically integrated result indicate rapid oscillations in susceptibility due to the interference between the Dirac cones.

%{\textcolor{red}{{\bf{Experimetal possibility/number?}}}} 
As far as practical realization of our results are concerned, it may be possible to deposit 
magnetic adatoms such as Co or Fe on bulk borophene to study RKKY exchange interaction in it. 
%might seem that depositing magnetic adatoms such as Co or Fe on borophene would be enough to study RKKY interaction therein. 
However, these adatoms consist of outermost {\it{s}} electrons, besides the {\it{d}} electrons
that act as essential magnetic moments, which can modify the exchange interaction. Moreover,
one has to be careful such that the band structure does not get modified significantly by
these impurities. Nevertheless, a simple molecule possessing localized magnetic moments
which interact via host material atoms and does not alter the bulk band structure, can be
deposited on 8-Pmmn borophene and may be a possible testbed for our theoretical predictions. %Kondo effect is expected to get suppressed compared to RKKY interaction in the undope case~\cite{uchoa2011kondo} but may play a competitive role for finite doping~\cite{allerdt2017competition} in case graphene.

Although we have focused on RKKY exchange interaction in 8-Pmmn Borophene as our host material,
qualitatively the results should be similar for any other 2D materials having tilted Dirac cone for \eg 
tilted graphene, organic conductor
$\alpha$-$({\rm {BEDT-TTF)_2}}$${\rm{ I_3}}$~\cite{goerbig2008tilted,katayama2006pressure} etc.
Then $\chi_{11}$ and $\chi_{12}$ would refer to the measure of 
RKKY exchange interaction (in units of $J\hbar^{2}/4$) corresponding to impurities being
placed on same sublattices and different sublattices respectively.

Here, we emphasize that 8-Pmmn borophene is ideally not a coplanar 2D Dirac material. Rather, it has a finite thickness due to the two kinds of atoms being non-coplanar. However, only those atoms 
located in a hexagonal manner in a buckled structure contribute to the formation of Dirac cones~\cite{lopez2016electronic}. Therefore, the finite thickness which we have neglected in our analysis, 
should manifests negligible effect on the RKKY exchange interaction.

%-----------------------------------------------
%{\it Acknowledgments.~}{.}
%-----------------------------------------------

%%%%%%%%%%%%%%%%%%%%%%%%%%%%%%%%%%%%%%%%%%%%%%%%%%%%%%

%-----------------------------------------------

%-----------------------------------------------
\bibliography{bibfile}{}
%-----------------------------------------------

\end{document}